\pdfoutput=1
\documentclass[a4paper,reprint,twocolumn,notitlepage,aip,nofootinbib]{revtex4-2}
\usepackage[left=1.5cm,right=1.5cm,top=1cm,bottom=1.5cm,includeheadfoot]{geometry}

\usepackage{tgtermes} % font
\usepackage{tgheros} % second font, used for sans-serif
\usepackage[T1]{fontenc}
\usepackage{amsmath}    % need for subequations
\usepackage{graphicx}  % needed for figures
\usepackage{bm}        % for math
\usepackage{amssymb}   % for math

\PassOptionsToPackage{hyphens}{url} % word breaks in url
\usepackage[breaklinks=true]{hyperref}

\usepackage{tikz}

\newcommand{\FF}{\mathbf{F}}
\newcommand{\rr}{\mathbf{r}}
\newcommand{\xx}{\mathbf{x}}
\newcommand{\jj}{\mathbf{j}}
\newcommand{\jmag}{\jj^{\rm mag}}
\renewcommand{\d}{\mathrm{d}} % for integrals

\renewcommand{\thesubsubsection}{\arabic{section}.\arabic{subsection}.\arabic{subsubsection}}
\usepackage{titlesec}
\titleformat{\subsubsection}[runin]{\sffamily\bfseries\small}{\thesubsubsection}{1em}{}

\begin{document}

\title{Local-density correlation functional from the force-balance equation}

\author{Nicolas Tancogne-Dejean}
  \email{nicolas.tancogne-dejean@mpsd.mpg.de}
 \affiliation{Max Planck Institute for the Structure and Dynamics of Matter and Center for Free-Electron Laser Science, Hamburg, Germany}
 \affiliation{European Theoretical Spectroscopy Facility (ETSF)}

\author{Markus Penz}
\affiliation{Max Planck Institute for the Structure and Dynamics of Matter and Center for Free-Electron Laser Science, Hamburg, Germany}
\affiliation{Department of Computer Science, Oslo Metropolitan University, Oslo, Norway}
 
\author{Michael Ruggenthaler}
\affiliation{Max Planck Institute for the Structure and Dynamics of Matter and Center for Free-Electron Laser Science, Hamburg, Germany}
\affiliation{The Hamburg Center for Ultrafast Imaging, Hamburg, Germany}
 
 \author{Angel Rubio}
  \email{angel.rubio@mpsd.mpg.de}
\affiliation{Max Planck Institute for the Structure and Dynamics of Matter and Center for Free-Electron Laser Science, Hamburg, Germany}
 \affiliation{European Theoretical Spectroscopy Facility (ETSF)}
 \affiliation{Nano-Bio Spectroscopy Group, Universidad del Pa\'is Vasco, San Sebasti\'an, Spain }
\affiliation{Center for Computational Quantum Physics (CCQ), The Flatiron Institute, New York NY 10010,~USA}

\begin{abstract}
The force-balance equation of time-dependent density-functional theory presents a promising route towards obtaining approximate functionals, however, so far, no practical correlation functionals have been derived this way. In this work, starting from a correlated wavefunction proposed originally by Colle and Salvetti [Theoret.\ Chim.\
Acta 37, 329 (1975)], we derive an analytical correlation-energy functional for the ground state based on the force-balance equation. 
The new functional is compared to the local-density correlation of the homogeneous electron gas and we find an increased performance for atomic systems, while it performs slightly worse on solids.
From this point onward, the new force-based correlation functional can be systematically improved.
\end{abstract}

\maketitle
%%%%%%%%%%%%%%%%%%%%%%%%%%%%%%%%%%%%%%%%%%%%%%%%%%%%%%%%%%%%%%%%%%%%%%%%%%%%%%%%%%%%%%%%%%%%%%%%%%%%%%%%%%%%%%%%%%%%%%%%%%%%%%%%%%
\section{Introduction}
%%%%%%%%%%%%%%%%%%%%%%%%%%%%%%%%%%%%%%%%%%%%%%%%%%%%%%%%%%%%%%%%%%%%%%%%%%%%%%%%%%%%%%%%%%%%%%%%%%%%%%%%%%%%%%%%%%%%%%%%%%%%%%%%%%
The quest for more accurate \textit{ab initio} descriptions of materials has been driving density-functional theories (DFT) towards ever improving energy functionals~\cite{toulouse2022review}. However, when it comes to describing excited states and dynamics of materials, we often have to rely on the adiabatic approximation.  The main bottleneck in obtaining time-dependent functionals beyond the adiabatic approximation is the lack of a clear path towards functionals not obtained as a functional derivative of the exchange-correlation energy. Of course, a lot of prior work explored how to go beyond the adiabatic approximation using various strategies like linear-response-based functionals~\cite{PhysRevLett.55.2850,PhysRevLett.79.1905}, current functionals~\cite{PhysRevLett.77.2037,PhysRevLett.79.4878}, time-dependent optimized effective potentials~\cite{PhysRevLett.100.056404,PhysRevLett.118.243001}, many-body-based functionals ~\cite{PhysRevB.72.235109}, and density-matrix coupled approximations~\cite{lacombe2019density}. Several exact constraints are now also known for the exchange-correlation potential of time-dependent density-functional theory (TDDFT), such as the harmonic potential theorem~\cite{PhysRevLett.73.2244,PhysRevLett.74.3233} and the zero-force theorem~\cite{PhysRevLett.74.3233,vignale1995sum}, among others~\cite{lacombe2023non}.

In this work, we demonstrate that it is possible to build an analytical exchange-correlation energy functional starting from the local force-balance equation approach and approximating the correlation force. This approach is very attractive, as there exists a clear link between the energy and the force at equilibrium, thanks to the virial relation, and also between the non-adiabatic exchange-correlation potential of TDDFT and the forces, thanks to a Sturm--Liouville-type equation~\cite{FBEx-2024}. The route via approximated exchange-correlation forces is therefore a potentially unifying approach that brings together equilibrium DFT and non-equilibrium TDDFT.
Here, we demonstrate that a simple ansatz for the correlated wavefunction, proposed originally by \citet{colle1975approximate} (which lies at the root of the widely used Lee--Yang--Parr (LYP) correlation energy functional~\cite{PhysRevB.37.785}) can be used for approximating the exchange-correlation force. With this, we recover as the exchange part the recently proposed local-exchange force and energy functional~\cite{FBEx-2024}, originally derived by \citet{PhysRevLett.62.489} in a different context, but we also obtain an approximated correlation force, which we show is connected to a correlation energy.

This work is organized as follow: In Sec.~\ref{sec:fbe} we present the basic force-balance equation for collinear-spin DFT. Then we show in Sec.~\ref{sec:functional} how to build an approximation for the exchange-correlation force density. Finally, in Sec.~\ref{sec:results} we perform simulations with our correlated local-density approximation and discuss its performance on different types of electronic systems, ranging from isolated atoms and molecules to periodic solids.
Our conclusions are presented in Sec.~\ref{sec:conclusions}.

%%%%%%%%%%%%%%%%%%%%%%%%%%%%%%%%%%%%%%%%%%%%%%%%%%%%%%%%%%%%%%%%%%%%%%%%%%%%%%%%%%%%%%%%%%%%%%%%%%%%%%%%%%%%%%%%%%%%%%%%%%%%%%%%%%
\section{Method}
\subsection{Force-balance equation}
\label{sec:fbe}
%%%%%%%%%%%%%%%%%%%%%%%%%%%%%%%%%%%%%%%%%%%%%%%%%%%%%%%%%%%%%%%%%%%%%%%%%%%%%%%%%%%%%%%%%%%%%%%%%%%%%%%%%%%%%%%%%%%%%%%%%%%%%%%%%%

To start with, we consider the $N$-electron non-relativistic Hamiltonian (in Hartree atomic units $e = \hbar = m_e = (4\pi\epsilon_0)^{-1} = 1$)
\begin{equation}
 \hat{H} =-\frac{1}{2}\sum_{k=1}^N \nabla^2_k + \sum_{k=1}^N v_{\rm ext}(\rr_k,t) + \sum_{k>l} \frac{1}{|\rr_k-\rr_l|},
\label{hamiltonian}
\end{equation}
where $v_{\rm ext}(\rr,t)$ is the time-dependent external one-particle potential acting on the electrons. We collect position and spin coordinates as $\xx_k =(\rr_k \sigma_k)$.
The spin-resolved equation of motion of the current density, also called ``local force-balance equation'', is~\cite{PhysRevLett.82.3863, stefanucci2013, tchenkoue2019force, FBEx-2024}
\begin{equation}\label{eq:force-balance}
\partial_t \jj(\xx,t) = -\rho(\xx,t)\nabla v_{\rm ext}(\rr,t) + \FF_T(\xx,t) + \FF_W(\xx,t). 
\end{equation}
This expression introduces the exact interaction-stress and momentum-stress force densities, respectively,
\begin{equation}\label{eq:interaction-force}
\begin{aligned}
\FF_W(\rr\sigma,t)=&- 2\sum_{\sigma'}\!\!\int\! (\nabla |\rr'-\rr|^{-1}) \\
&\times \rho^{(2)}(\rr\sigma,\rr'\sigma',\rr\sigma,\rr'\sigma',t) \,\d \rr',\\
\end{aligned}
\end{equation}
\begin{equation}\label{eq:momentum-stress-force}
\FF_T(\rr\sigma,t) = \left.\frac{1}{4} (\nabla - \nabla')(\nabla^2 - \nabla'^2)\rho^{(1)}(\rr\sigma,\rr'\sigma,t)\right|_{\rr'=\rr}. 
\end{equation}
Here, the $p^{\rm th}$-order reduced density matrix is given by
\begin{equation}
\begin{aligned}
\rho^{(p)}&(\xx_1, \ldots, \xx_p,\xx_1', \ldots, \xx_p', t)\\
=& \frac{N!}{p!(N-p)!} \sum_{\sigma_{p+1} \ldots \sigma_N}\int \Psi(\xx_1,\ldots,\xx_p,\xx_{p+1},\ldots,\xx_N,t)\\
& \quad \Psi^{*}(\xx'_1,\ldots,\xx_p',\xx_{p+1},\ldots,\xx_N,t) \,\d\rr_{p+1}\ldots\d\rr_N.
\label{p-RDM}
\end{aligned}
\end{equation}
We indicate the force densities coming from the solution $\Psi$ of the fully interacting problem as $\FF_W[\Psi]$ and $\FF_T[\Psi]$. The auxiliary non-interacting Kohn--Sham (KS) problem (Eq.~\eqref{hamiltonian} without interaction term and with a different spin-resolved external potential $v_{\rm s}(\xx,t)$) has a Slater-determinant solution $\Phi$ and only includes $\FF_T[\Phi]$. In analogy to Eq.~\eqref{eq:force-balance} we have for the KS system
\begin{equation}\label{eq:force-balance-s}
\partial_t \jj_{\rm s}(\xx,t) = -\rho_{\rm s}(\xx,t)\nabla v_{\rm s}(\xx,t) + \FF_T([\Phi],\xx,t).
\end{equation}

Considering first the static ground-state setting, we impose that both systems generate the same ground-state density, $\rho(\xx) = \rho_{\rm s}(\xx)$. Since $\partial_t \jj(\xx) = \partial_t \jj_{\rm s}(\xx) = 0$ for any eigenstate, by subtracting Eq.~\eqref{eq:force-balance-s} from Eq.~\eqref{eq:force-balance} we find with the definition of the Hartree-exchange-correlation (Hxc) potential $v_{\rm Hxc}(\xx) = v_{\rm s}(\xx) - v_{\rm ext}(\rr)$ that
\begin{equation}\label{eq-Hxc-potential}
\rho\nabla v_{\rm Hxc} = -\FF_{\rm Hxc}[\Phi, \Psi] = \FF_T[\Phi] - \FF_T[\Psi] - \FF_W[\Psi],
\end{equation}
which defines $\FF_{\rm Hxc}$. This expression has been previously derived based on a differential virial relation~\cite{PhysRevA.51.2040}.
The Hxc force density can be partitioned in analogy to the usual partition of the energy into a Hartree-exchange (Hx) force density and a correlation force density,
\begin{align}\label{eq:Hxcforces}
& \FF_{\rm Hx}[\Phi] = \FF_W[\Phi],\\
& \FF_{\rm c}[\Phi,\Psi] = \FF_T[\Psi] -\FF_T[\Phi]  + \FF_W[\Psi] - \FF_W[\Phi].
\end{align}
While the exchange part can be directly turned into a local-exchange potential~\cite{FBEx-2024},
the correlation part can be split again into a kinetic-correlation contribution $\FF_T[\Psi] -\FF_T[\Phi]$ and an interaction-correlation contribution $\FF_W[\Psi] - \FF_W[\Phi]$.
As we will show below, thanks to the local force-balance equation Eq.~\eqref{eq-Hxc-potential}, an approximation for the correlation force density can be turned into an exchange-correlation potential.

In the time-dependent case, only the longitudinal part of the current is guaranteed to be the same between the two systems, thanks to the continuity equation,
\begin{equation}
    \partial_t\rho(\xx,t) = \partial_t\rho_{\rm s}(\xx,t) = -\nabla\cdot\jj(\xx,t) = -\nabla\cdot\jj_{\rm s}(\xx,t).
\end{equation}
Hence, the divergence of Eq.~\eqref{eq-Hxc-potential} still holds at all times and forms the fundamental equation of TDDFT~\cite{ruggenthaler2015b}
\begin{equation}\label{eq:ExchangeTimeDependent}
    \nabla \cdot \left[\rho(\xx,t) \nabla v_{\rm Hxc}(\xx,t) \right] = - \nabla \cdot \FF_{\rm Hxc}(\xx,t).
\end{equation}
The local exchange-correlation potential in TDDFT is now determined from the exchange-correlation force density by inverting a Sturm--Liouville-type equation. The local  exchange-correlation potential in TDDFT obtained from $\FF_{\rm Hxc}$ now allows to go beyond the adiabatic approximation~\cite{fuks2018exploring}.

Note that while clearly Eq.~\eqref{eq-Hxc-potential} implies Eq.~\eqref{eq:ExchangeTimeDependent}, this does not generally hold in the other direction since any transverse vector-field contribution in Eq.~\eqref{eq-Hxc-potential} gets canceled by the application of the divergence.
On the other hand, it should be noted that Eqs.~\eqref{eq-Hxc-potential} and \eqref{eq:ExchangeTimeDependent} are indeed equivalent if (i) $\rho$ is spatially uniform, (ii) the spatial domain is one-dimensional, or (iii) the vector field $\FF_{\rm Hxc}/\rho$ is purely longitudinal anyway (because it is then possible to find a $v_{\rm Hxc}$ such that $\rho\nabla v_{\rm Hxc} = -\FF_{\rm Hxc}$ as in Eq.~\eqref{eq:ExchangeTimeDependent}). In all those situation, the adiabatic approximation from Eq.~\eqref{eq-Hxc-potential} is thus sufficient also for the time-dependent setting.

%%%%%%%%%%%%%%%%%%%%%%%%%%%%%%%%%%%%%%%%%%%%%%%%%%%%%%%%%%%%%%%%%%%%%%%%%%%%%%%%%%%%%%%%%%%%%%%%%%%%%%%%%%%%%%%%%%%%%%%%%%%%%%%%%%
\subsection{Deriving a local-density correlation functional}
\label{sec:functional}
%%%%%%%%%%%%%%%%%%%%%%%%%%%%%%%%%%%%%%%%%%%%%%%%%%%%%%%%%%%%%%%%%%%%%%%%%%%%%%%%%%%%%%%%%%%%%%%%%%%%%%%%%%%%%%%%%%%%%%%%%%%%%%%%%%

\subsubsection{Correlation force}

Our aim is to find an expression for the correlation force density, to be used in the force-balance equation. In the following, we make use of the approximation proposed originally by \citet{colle1975approximate} for the correlation energy to find an expression for the correlation force density.
%
%In order to find an approximation for the force density, we must therefore find approximations to $\rho^{(2)}$ and $\rho^{(1)}$. 
% Previously, we used the local Hartree approximation, for which 
% \begin{equation}
%  \rho^{(2)}(\rr,\rr',\rr,\rr')\approx \rho^{(2)}_{\rm s}(\rr,\rr') =  \frac{1}{2}\left(\rho(\rr) \rho(\rr') -  \frac{1}{2}\rho^{(1)}_{\rm s}(\rr,\rr')\rho^{(1)}_{\rm s}(\rr',\rr) \right)
% \end{equation}
% and $\rho^{(1)}\approx \rho^{(1)}_{\rm s} = \sum_{k,\sigma} \varphi_{k}(\rr \sigma)\varphi^{*}_{k}(\rr' \sigma)$.
% 
%% already mentioned before
%In this work, we go beyond the local Hartree-exchange approximation that we recently proposed~\cite{FBEx-2024} and take correlation force densities into account as well. More precisely, we consider an approximation for the two-body reduced density matrix (2RDM)~\cite{colle1975approximate}
They start with an approximation for the two-body reduced density matrix (2RDM),
\begin{equation}
\begin{aligned}
 \rho^{(2)}&(\xx_1,\xx_2,\xx_1', \xx_2') \approx \rho^{(2)}_{\rm s}(\xx_1,\xx_2,\xx_1', \xx_2')
 \\
 &\times
 \left( 1 - \nu(\rr_1,\rr_2)  - \nu(\rr_1',\rr_2') + \nu(\rr_1,\rr_2)\nu(\rr_1',\rr_2')\right),
 \label{eq:2rdm_CS}
\end{aligned}
\end{equation}
together with the 1RDMs set equal,
\begin{equation}
 \rho^{(1)}(\xx,\xx')\approx \rho^{(1)}_{\rm s}(\xx,\xx') = \sum_{k} \varphi_{k}(\xx)\varphi^{*}_{k}(\xx').    
\end{equation}
Here, the diagonal part of the single Slater determinant 2RDM is given by
\begin{equation}
\begin{aligned}
\rho_{\rm s}^{(2)}&(\xx,\xx',\xx,\xx') \\
&= \frac{1}{2}\left(\rho(\xx) \rho(\xx') - \delta_{\sigma\sigma'} |\rho_{\rm s}^{(1)}(\xx,\xx')|^2 \right).
\end{aligned}
\end{equation}
In Eq.~\eqref{eq:2rdm_CS} the spinless correlation factor $\nu$ is defined in the inter-particle frame as
\begin{equation}
 \nu(\rr,\mathbf{s}) = e^{-\beta^2(\rr)s^2}\left(1-\Theta(\rr)\left(1+\frac{s}{2} \right)\right),
\end{equation}
with $s=|\rr'-\rr|$, $\rr=\frac{1}{2}(\rr'+\rr)$, and $\beta(\rr)=q\rho^{1/3}(\rr)$, with $q>0$.
As remarked by \citet{colle1975approximate}, the volume on which the correlation function $\nu$ is appreciably different from zero is dictated by $\beta$. This (local) volume can be defined as
\begin{equation}
 V = 4\pi \int \d r\, e^{-\beta^2r^2} r^2 = \frac{\pi^{3/2}}{\beta^3}.
\end{equation}
A possible choice is $V = q/\rho$, the volume of exclusion in Wigner's formula, which leads to $\beta(\rr) = q\rho(\rr)^{1/3}$. This allows for defining $\beta$ from the local value of the density.
Colle and Salvetti showed that the function $\Theta$ is well approximated by
\begin{equation}
 \Theta(\rr) \approx \frac{\sqrt{\pi}\beta(\rr)}{1+\sqrt{\pi}\beta(\rr)}.
\end{equation}
In the spirit of the original work of Colle and Salvetti, the parameter $q$ is left as a free parameter for the moment. We will explain how we determine this parameter later.

Using the ansatz from Eq.~\eqref{eq:2rdm_CS} for the 2RDM, we obtain 
\begin{widetext}
\begin{equation}
\FF_W^{\rm CS}(\rr\sigma)=- 2\sum_{\sigma'}\int\d \rr'\, (\nabla |\rr'-\rr|^{-1}) \rho^{(2)}_{\rm s}(\xx,\xx',\xx,\xx')  \left( 1 - 2\nu(\rr,\rr')  + \nu^2(\rr,\rr')\right) = \FF_{\rm H} + \FF_{\rm x} + \FF_{\rm c}^{\rm CS}.
\end{equation}
It is important to note that this force density respects the zero-force theorem, as it is straightforward to show that $\int\d\rr\, \FF_W^{\rm CS}(\rr\sigma) = 0$ by symmetry of $\rr$ and $\rr'$.
Using the above expression, the correlation force reads
\begin{equation}
\FF_{\rm c}^{\rm CS}(\rr\sigma)=- 2\int\d \rr' \, (\nabla |\rr'-\rr|^{-1}) \rho^{(2)}_{\rm s}(\xx,\xx',\xx,\xx') \left( \nu^2(\rr,\rr') - 2\nu(\rr,\rr')   \right),
\end{equation}
or, in the interparticle coordinate frame (with $\mathbf{s}=\rr'-\rr$ the interparticle coordinate and its normalized version $\mathbf{\hat{s}}=\mathbf{s}/s$),
\begin{equation}
\begin{aligned}
\FF_{\rm c}^{\rm CS}(\rr\sigma)= 2\sum_{\sigma'}\int \d \mathbf{s}\, \frac{\mathbf{\hat{s}}e^{-\beta^2(\rr)s^2}}{s^2} \rho^{(2)}_{\rm s}(\rr\sigma,(\rr+\mathbf{s})\sigma',\rr\sigma,(\rr+\mathbf{s})\sigma') \Big( & e^{-\beta^2(\rr)s^2}\left( 1-\Theta\left(\rr+\frac{\mathbf{s}}{2}\right)\left(1+\frac{s}{2} \right)\right)^2
\\
& - 2\left(1-\Theta\left(\rr+\frac{\mathbf{s}}{2}\right)\left(1+\frac{s}{2} \right)\right)\Big).
\end{aligned}
\end{equation}
\end{widetext}
We have thus arrived at an approximate expression for the correlation force density, which here only contains interaction correlations and no kinetic correlations. 
In fact, we have simply defined the exchange and correlation force density that correspond to Hartree plus exchange, together with the interaction correlations treated at the same level of approximation as used by Colle and Salvetti for deriving their correlation energy ($E_{\rm c}^{\rm CS}$; see Eq. (9) in Ref.~\onlinecite{colle1975approximate}). Our force should therefore be related to their energy. To see this, we show in App.~\ref{app:virial} that $ \FF_W^{\rm CS}(\rr\sigma)$ indeed fulfills the virial relation
\begin{equation}
 E_{\rm c}^{\rm CS} = \sum_\sigma\int \d\rr \, \rr \cdot \FF_{\rm c}^{\rm CS}(\rr\sigma).
 \label{eq:virial}
\end{equation}
This shows that our force density is the correct force density associated with the Colle--Salvetti correlation energy, before performing any gradient expansion. We note that this virial relation holds as in the case of exchange~\cite{FBEx-2024}, since we only consider interaction correlations here. 
Indeed, since the 1RDM of the KS system and of the fully interacting one are equal, no kinetic correlations are included in this work. We also note that this approach is similar to the density-matrix coupled approximations proposed by Lacombe \textit{et al.}~\cite{lacombe2019density}, with the difference that we are not suggesting here to have an equation of motion for the 2RDM.

We could of course stop here and follow the work of Colle and Salvetti in approximating the energy and then assume functional differentiability to get the corresponding exchange-correlation potential. However, this approach has the drawback that it does not allow us to explore the time-dependent case, where one needs to solve the Sturm--Liouville Eq.~\eqref{eq:ExchangeTimeDependent}, and it would thus confine us to the adiabatic approximation. As our goal is to find an approximation for the correlation force density, we show how we can further approximate the expression of the force density, in order to get a numerically simpler expression for the correlation force density.

For this we perform a gradient expansion, see App.~\ref{app:force_density} for more details.
The zero-order term vanishes by symmetry, and in the following we consider only the first order term of the gradient expansion. Going to the next non-zero order (third order) would lead to a semi-local meta-GGA functional, and we delegate this derivation to future work. The spin-polarized and spin-unpolarized cases are presented separately, as their derivations show subtle differences.
% We can now expand the two-body reduced density matrix and its gradient up to second order in $s$.  We want here to perform a Taylor expansion using a relation like the one used for scalar quantities
% \begin{equation}
%  \int f(s) F(\rr,\mathbf{s}) \d\mathbf{s} = 4\pi  \int f(s) s^2 \Big[\left(1 + \frac{1}{3!}s^2\nabla'^2 + \ldots \right) F(\rr,\rr')\Big]_{\rr'=\rr} \d s 
% \end{equation}
% which can be proven using the fact that $f(\rr,\rr+\mathbf{s}) = e^{\mathbf{s}\cdot\nabla'} f(\rr,\rr')|_{\rr'=\rr}$.
% Before using this, we need to generalize this result for the vector-field case we are interested in.
%    
After some simple algebra and using that $\beta(\rr) = q\rho(\rr)^{1/3}$, we arrive at the expression for the first-order gradient expansion of the correlation force density in the spin-unpolarized case
\begin{widetext}
\begin{equation}
\FF_{\rm c}^{\rm CS}(\rr)= \frac{\pi}{3q^2} \rho(\rr) \left\{3\left[\nabla \left(\Theta^2(\rr)-1\right) \rho^{1/3}(\rr)\right] 
 - 5\rho^{1/3}(\rr)\Theta(\rr)\left[\nabla \Theta(\rr)\right]\right\}.
 \label{eq:correlation_force_unpol}
\end{equation}
(Here and in the following the notation $[\nabla\ldots]$ means that the gradient is applied to the whole expression inside the square brackets.)
In the spin-polarized case, the correlation force density is found to be
\begin{equation}\label{eq:correlation_force_pol}
\begin{aligned}
\FF_{\rm c}^{\rm CS}(\rr\sigma) =\;& \frac{\pi}{3 q^2 \rho^{2/3}(\rr)}
\rho_\sigma(\rr)\Big\{\left(\Theta^2(\rr)-1\right) [\nabla \rho(\rr)] 
 + \rho(\rr)\Theta(\rr)\left[\nabla \Theta(\rr)\right]\Big\}\\
 &- \frac{\pi}{3 q^2 \rho^{2/3}(\rr)}
\rho_\sigma(\rr)\Big\{\left(\Theta^2(\rr)-1\right) [\nabla \rho(\rr)\zeta_\sigma(\rr)] 
 + \rho(\rr)\zeta_\sigma(\rr)\Theta(\rr)\left[\nabla \Theta(\rr)\right]\Big\}.
\end{aligned}
\end{equation}
\end{widetext}
Here, we write $\rho_\sigma(\rr) = \rho(\rr\sigma)$ and let $\rho_{\bar{\sigma}}$ describe the density of the opposite spin channel. We further introduced the total (spin-summed) density $\rho=\rho_\sigma+\rho_{\bar{\sigma}}$ and the relative spin polarization $\zeta_\sigma=(\rho_\sigma-\rho_{\bar{\sigma}})/\rho$. 
%
% \begin{eqnarray}
% \FF_{\rm c}^{\rm CS}(\rr\sigma)= \frac{2\pi}{3 q^2 \rho^{2/3}(\rr)}
% \rho_\sigma(\rr)\Bigg[\Big(\Phi^2(\rr)-1\Big) [\nabla \rho_{\bar{\sigma}}(\rr)] 
%  + \rho_{\bar{\sigma}}(\rr)\Phi(\rr)\Big[\nabla \Phi(\rr)\Big]\Bigg]
% \end{eqnarray}
%simply obtained by multiplying the unpolarized force density by a factor $\frac{1-\zeta^2(\rr)}{2}$ for the force density,
% with $\beta$ been computed from the total (spin-summed) density.
% \begin{eqnarray}
% \FF_{\rm c}^{\rm CS}(\rr\sigma)= \frac{\pi}{3q^2} \rho(\rr) \frac{1-\zeta^2(\rr)}{2} \Bigg(\Phi^2(\rr)-1\Big) \rho^{1/3}(\rr)] 
%  - 5\rho^{1/3}(\rr)\Phi(\rr)\Big[\nabla \Phi(\rr)\Big]\Bigg) 
% \end{eqnarray}
% where we introduce the relative spin polarization $\zeta=(\rho_\uparrow-\rho_\downarrow)/\rho$. 
%
% To further simplify this, we rewrite the term in bracket in terms of the total (spin-summed) density and in terms of a spin-difference term, in a relative spin polarization $\zeta_\sigma=\frac{\rho_\sigma-\rho_{\bar{\sigma}}}{\rho}$. In particular, we have $\rho_{\bar{\alpha}} = \rho\frac{1-\zeta_\alpha}{2}$
% Simple manipulations of the expression leads to 
% \begin{eqnarray}
% \FF_{\rm c}^{\rm CS}(\rr\sigma)= \frac{\pi}{3 q^2}
% \rho_\sigma(\rr)(1-\zeta_\sigma(\rr))\Bigg(3[\nabla \Big(\Phi^2(\rr)-1\Big) \rho^{1/3}(\rr)] 
%  - 5\rho^{1/3}(\rr)\Phi(\rr)\Big[\nabla \Phi(\rr)\Big]\Bigg) 
%  \label{eq:correlation_force_pol}
% \end{eqnarray}
% where in order to derive this expression, we made the usual approximation~\cite{PhysRevLett.77.3865} that consists in neglecting the small corrections coming from $\nabla\zeta_\sigma$.
%
The first term of the force is the spin-averaged force density, and the second term,  associated with $\zeta$, corresponds to the contribution from the spin polarization. In the unpolarized limit ($\rho_\sigma = \rho_{\bar{\sigma}} = \rho/2$ and $\zeta=0$) we recover the unpolarized case.  

\subsubsection{Correlation potential}

The above expression for the spin-unpolarized case (Eq.~\eqref{eq:correlation_force_unpol}) does not allow to tell by simple inspection what the corresponding correlation potential is. In order to find the potential associated with our correlation force density, we first rewrite the second term in the force density Eq.~\eqref{eq:correlation_force_unpol} as a gradient using the relation $\nabla [f(\beta)] = [\nabla \beta]f'(\beta)$. Performing the integration over $\beta$ leads to
\begin{equation}
\begin{aligned}
 \rho^{1/3}\Theta[\nabla \Theta] 
 &= [\nabla\beta]\frac{ \pi \beta^2 }{q(1+\sqrt{\pi}\beta)^3}\\
 &= \frac{1}{q\sqrt{\pi}}\nabla \left[ \ln(1+\sqrt{\pi}\beta) 
 -\frac{1}{2}\frac{\Theta^2}{\pi\beta^2} 
 + \frac{2\Theta}{\sqrt{\pi}\beta}\right].
\end{aligned}
\end{equation}
This allows to express the entire curly bracket of Eq.~\eqref{eq:correlation_force_unpol} as a gradient of a scalar field, showing that in the spin-unpolarized case the correlation force density is associated to a purely longitudinal correlation vector field, and we can now directly read off the correlation potential from the relation $\mathbf{F}_{\rm c} = -\rho\nabla v_{\rm c}$.
This also implies that both Eq.~\eqref{eq-Hxc-potential} and the Sturm--Liouville equation for the time-dependent case Eq.~\eqref{eq:ExchangeTimeDependent} lead to the same potential. Consequently the potential for the ground-state and time-dependent cases are identical in the unpolarized case, even in the presence of a finite current in our system. This implies that this correlation functional will \emph{not} contain any memory effect, even though we do not rely on the adiabatic approximation. Indeed, memory requires the correlation vector field ($\mathbf{F}_{\rm c}/\rho$) to acquire a \emph{transverse} part when the system gets excited out of equilibrium, as was noted at the end of Sec.~\ref{sec:fbe}. 

The correlation potential for the unpolarized case is thus given by
\begin{equation}\label{eq:vc-unpolarized}
\begin{aligned}
 v_{\rm c}(\rr) = \;&\frac{\pi}{q^3}\left(\Theta^2-1\right) \beta \\
 &- \frac{5\sqrt{\pi}}{3q^3}\left[ \ln(1+\sqrt{\pi}\beta) 
 -\frac{1}{2}\frac{\Theta^2}{\pi\beta^2} 
 + \frac{2\Theta}{\sqrt{\pi}\beta}\right].
\end{aligned}
\end{equation}
This determines the potential up to a constant. In the limit $\beta\to0$ we have $v_{\rm c}(\rr) \to -5\sqrt{\pi}/(2q^3)$. This constant term is not relevant in the Hamiltonian, but it must be added to our potential in order to get the correct limit for the correlation energy.

Our potential only depends on the local value of the density and can therefore be considered a local-density approximation to the correlation potential.
The corresponding force density is purely longitudinal, and hence does not contain memory effects. This is in agreement with the harmonic theorem of Dobson and Vignale~\cite{PhysRevLett.73.2244,PhysRevLett.74.3233}, which states that the local-density functionals cannot have memory. 

We now consider the spin-polarized case. If the force density would be longitudinal, we would be able to define a scalar potential $v(\beta,\zeta_\sigma)$ such that 
\begin{equation}
 \FF_{\rm c}^{\rm CS}(\rr\sigma) = \rho_\sigma(\rr)\left[ [\nabla \beta]\frac{\partial v(\beta,\zeta_\sigma)}{\partial \beta} + [\nabla \zeta_\sigma]\frac{\partial v(\beta,\zeta_\sigma)}{\partial \zeta_\sigma}\right].
\end{equation}
We can rewrite the spin-polarized part of Eq.~\eqref{eq:correlation_force_pol} (second line) as
\begin{equation}
\begin{aligned}
&\FF_{\rm c,pol}^{\rm CS}(\rr\sigma) = 
 -
 \frac{\pi}{3 q^3 }
\rho_\sigma(\rr)\bigg[
  [\nabla \zeta_\sigma(\rr)]\Big(\Theta^2(\rr)-1\Big) \beta(\rr) \\
&+ [\nabla \beta(\rr)] \left(3\zeta_\sigma(\rr) \Big(\Theta^2(\rr)-1\Big) 
 + \zeta_\sigma(\rr)\frac{ \pi \beta^2 }{(1+\sqrt{\pi}\beta)^3}\right)\bigg].
\end{aligned}
\end{equation}
This allows to identify $\partial v(\beta,\zeta_\sigma)/\partial \beta$ and $\partial v(\beta,\zeta_\sigma)/\partial \zeta_\sigma$. 
Using these expressions, one can compute the mixed derivative $\partial^2 v(\beta,\zeta_\sigma)/\partial \beta \partial \zeta_\sigma$ from both expressions. It is simple to show that this leads to a violation of the symmetry of second derivatives while the partial derivatives are continuous functions here. This shows that the force is \emph{not} a conservative force. The same can also be seen by computing the curl of the spin-polarized part of the force. This term can be shown to be proportional to $\nabla \zeta_\sigma \times \nabla \rho$, or equivalently to $\nabla m_z \times \nabla \rho$, where $m_z=\rho_\uparrow-\rho_\downarrow$. The force vector field is then purely longitudinal only when the system is fully polarized ($\zeta_\sigma=1$) or when the system is non-magnetic ($\zeta_\sigma=0$). 
The implications of this result are interesting: The spin-polarized force density, unlike the spin-unpolarized force density, carries some memory, as solving Eqs.~\eqref{eq-Hxc-potential} and \eqref{eq:ExchangeTimeDependent} will not lead to the same result. This result might sound unexpected, but one needs to remember that in the spin-polarized case the current density also contains a magnetization current of the form $\jmag = c\nabla \times ( m_z \mathbf{\hat{e}_z}) = c[\nabla m_z]\times \mathbf{\hat{e}_z}$. Our force density, that is a functional of the magnetization density, is thus related to a non-trivial part of the transverse current, which is then expected to contain memory.
This result reveals an interesting aspect of memory in TDDFT: there is a contribution to the memory from the spins, due to the correlation between electrons that couples the different spin channels. 

Let us comment on how to proceed with using our spin-polarized correlation force density in order to get the corresponding correlation potential and the correlation energy. As already discussed, the energy is obtained via the virial relation Eq.~\eqref{eq:virial}. For the potential, we are left with two choices. One is to solve Eq.~\eqref{eq-Hxc-potential} assuming no time-dependent currents (either in the ground state or invoking the adiabatic approximation) using the Poisson equation $-\Delta v_\mathrm{Hxc} = \nabla\cdot\FF_\mathrm{Hxc}/\rho$, as done in Ref.~\onlinecite{FBEx-2024}. The other option is to solve the Sturm--Liouville Eq.~\eqref{eq:ExchangeTimeDependent}.
Yet, if we neglect terms associated with $\nabla \zeta_\sigma$, as it is common in spin-polarized energy functionals~\cite{PhysRevB.46.6671}, then we even get a conservative force and obtain the corresponding potential directly from the local force-balance equation.
Indeed, this approximation gives us
\begin{eqnarray}
\FF_{\rm c}^{\rm CS}(\rr\sigma)&=& \rho_\sigma(\rr)[\nabla  (1-\zeta_\sigma(\rr)) v_{\rm c}(\rr)],
\label{eq:approx_polarized_force}
\end{eqnarray}
and thus
\begin{eqnarray}
 v_{\rm c}(\rr\sigma)  = (1-\zeta_{\sigma})v_{\rm c}(\rr),
 \label{eq:polarized_potential_approx}
\end{eqnarray}
where $ v_{\rm c}(\rr)$ without the spin coordinate is from the spin-unpolarized case Eq.~\eqref{eq:vc-unpolarized}. However, from this potential one cannot define a correlation energy from an energy density, i.e., of the form $E_{\rm c} = \int \d\rr\, \epsilon_{\rm c}[\rho,\zeta_\uparrow-\zeta_\downarrow]$, as this leads again to a violation of the symmetry of second derivatives. For the spin-polarized case and under the approximation of neglecting $\nabla \zeta_\sigma$, we can only have a potential functional, whereas we also have an analytical energy functional for the unpolarized case, as discussed in the next section.

\subsubsection{Correlation energy}

We now look at the energy density corresponding to the above potential for the spin-unpolarized case. As the potential is a functional of the density only, it is straightforward to obtain the energy density by integration. We find after some algebra that for the unpolarized case
\begin{widetext}
\begin{equation}
 q^6\epsilon_{\rm c}[\rho] =  
 -\frac{4}{3} q^2\rho^{-1/3} 
 + \frac{19}{18}q^3\sqrt{\pi} 
 + \frac{13q}{6\sqrt{\pi}} \rho^{-2/3} 
 + \frac{1}{2\pi\rho(\sqrt{\pi}q\rho^{1/3} + 1)} 
 - \frac{5}{3} \ln(\sqrt{\pi}q\rho^{1/3} + 1)\Big(\frac{1}{\pi\rho} +q^3\sqrt{\pi}\Big) 
 -\frac{1}{2\pi\rho },
 \label{eq:correlation_energy}
\end{equation}
\end{widetext}
where the last constant is the integration constant, which we set such that $\rho\epsilon_{\rm c}[\rho] \to 0$ as $\rho\to 0$.

So far, we have not discussed how to select the value of the parameter $q$. For fixing it, we consider an electron gas with a uniform density $\rho = 3/(4\pi r_s^3)$, where $r_s$ is its Wigner--Seitz radius. The low- and high-density limits for the correlation energy per particle are known for this system. In particular, in the high-density limit, corresponding to $r_s\to 0$, it is known that the correlation energy density $\epsilon_{\rm c}$ of the unpolarized homogeneous electron gas scales as~\cite{toulouse2022review}
\begin{equation}
 \epsilon_{\rm c}(r_s) = A \ln r_s + B + C r_s \ln r_s + \mathcal{O}(r_s),
\end{equation}
with $A = (1-\ln 2)/\pi^2$, $B = -0.046921$, $C=0.009229$.
From the above expression of our correlation functional (Eq.~\eqref{eq:correlation_energy}), the leading order in $r_s$ is given by
\begin{equation}
 \epsilon_{\rm c}(r_s) \approx \frac{5}{3q^3}\sqrt{\pi} \ln r_s ,
\end{equation}
which allows us to define the parameter $q\approx 4.5631$ such that the leading order for $r_s\to 0$ is the one of the homogeneous electron gas. We find that the constant term is then
\begin{equation}
 \frac{19\sqrt{\pi}}{18q^3}-\frac{5}{3q^3}\ln\left(q\left(\frac{3\sqrt{\pi}}{4}\right)^{1/3}\right)\approx -0.08602,
\end{equation}
which is twice the value of $B$ for the homogeneous electron gas. The next order does not have the right scaling, as it scales as $r_s^3 \ln r_s$. In the polarized case, we use the same value of $q$ as in the unpolarized case.

%%%%%%%%%%%%%%%%%%%%%%%%%%%%%%%%%%%%%%%%%%%%%%%%%%%%%%%%%%%%%%%%%%%%%%%%%%%%%%%%%%%%%%%%%%%%%%%%%%%%%%%%%%%%%%%%%%%%%%%%%%%%%%%%%
\section{Results}
\label{sec:results}
%%%%%%%%%%%%%%%%%%%%%%%%%%%%%%%%%%%%%%%%%%%%%%%%%%%%%%%%%%%%%%%%%%%%%%%%%%%%%%%%%%%%%%%%%%%%%%%%%%%%%%%%%%%%%%%%%%%%%%%%%%%%%%%%%%
The polarized and unpolarized correlation functionals were implemented in the real-space code Octopus~\cite{tancogne2020octopus}.
\subsection{Homogeneous electron gas}
Before investigating the performance of our functional on periodic and finite systems, it is instructive to compare its performances compared to other known functionals for the case of the homogeneous electron gas, which is one of the cornerstone of deriving energy functionals in DFT, especially for periodic systems.

In order to assess our functional, we consider the correlation functionals of Wigner~\cite{TF9383400678}, Perdew--Zunger~\cite{perdew1981} and Hedin--Lundqvist~\cite{Hedin_1971}, which are all local-density approximations, as well as Lee--Yang--Parr (LYP) MGGA~\cite{PhysRevB.37.785} that is based on the same ansatz for the correlated wavefunction from Colle and Salvetti.
\begin{figure}[ht]
  \begin{center}
  \includegraphics[width=\columnwidth]{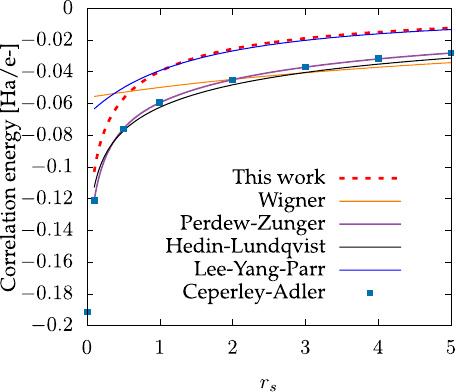}
    \caption{\label{fig:HEG} Correlation energy per electron for the homogeneous electron gas as a function of the Wigner--Seitz radius $r_s$ for different functionals, see main text for details. The dots represent the quantum Monte-Carlo results of \citet{PhysRevLett.45.566}.}
  \end{center}
\end{figure}

From Fig.~\ref{fig:HEG} it is clear that overall the correlation energy obtained from our functional underestimates the Perdew--Zunger and Hedin--Lundqvist functionals which are obtained by parameterizing the data of a quantum Monte-Carlo simulations~\cite{PhysRevLett.45.566}. Interestingly, even though we impose the high-density limit to be fulfilled, our functional converges to the LYP MGGA correlation energy functional in the low-density limit.

The reasons for the disagreement are clear: our approach and the LYP correlation share in common that they are based on a short-range expansion. Due to this, they completely miss the long-range correlations~\cite{PhysRevA.63.032513}. Moreover, the kinetic correlations are absent from these two functionals.
It is therefore expected that the proposed functional will not perform as good as LDA-based ones on the homogeneous electron gas, as long-range correlations are important for extended systems (see the next section for results). However, the force-based functional might provide interesting results for isolated systems, since the LYP correlation is known to improve upon LDA for finite systems.

\subsection{Periodic systems}

In order to assess the performance of the proposed functional, we considered a set of typical bulk semiconductors and insulators and we computed for each of them the band gap, using the experimental equilibrium position, but also the lattice constant and the bulk moduli, obtained from a fit by the Birch--Murnaghan equation of state~\cite{PhysRev.71.809} of seven self-consistent calculations performed at 0.94, 0.96, 0.98, 1, 1.02, 1.04, and 1.06 times the experimental equilibrium position. Our results are calculated for exchange-correlation LDA (LDA$_{\rm xc}$) using the modified Perdew--Zunger correlation~\cite{perdew1981}, but also LDA exchange and our proposed correlation functional (LDA$_{\rm x}$+FBE$_{\rm c}^{\rm CS}$) in order to evaluate the performance of the correlation part alone.
We employed PseudoDojo LDA pseudopotentials~\cite{van2018pseudodojo}. The simulation parameters for the different solids are given in App.~\ref{app:parameters}.
The results for the band gap are shown in Tab.~\ref{tab:bangaps}. The results for the lattice constant and the bulk moduli are shown in Tab.~\ref{tab:lattice_constants} and Tab.~\ref{tab:bulk_moduli}, respectively.

\begin{table}
\begin{ruledtabular}
\begin{tabular}{lcccc}
Approx. & LDA$_{\rm xc}$ & LDA$_{\rm x}$+FBE$_{\rm c}^{\rm CS}$  & Exp. \\
\hline
AlAs & 1.336 & 1.325 & 2.15~\cite{PhysRevMaterials.3.064603} \\
AlSb & 1.146 & 1.132 & 1.62~\cite{PhysRevMaterials.3.064603}\\
GaAs & 0.290 & 0.2874 & 1.52~\cite{PhysRevLett.102.226401}\\
ZnO & 2.604 & 0.872 & 3.44~\cite{Agapito_PRX}\\
AlP & 1.444 & 1.434 & 2.45~\cite{PhysRevLett.102.226401}\\
BN & 4.101 & 4.097  & 6.25~\cite{PhysRevLett.102.226401}\\
TiO$_2$ & 1.831 & 1.831 & 3.03-3.3~\cite{Agapito_PRX}\\
Ar &8.168 & 8.130 & 14.20~\cite{PhysRevLett.102.226401} \\
C &4.143 & 4.143 & 5.48~\cite{PhysRevLett.102.226401}\\
Si & 0.524 & 0.517 & 1.17~\cite{PhysRevLett.102.226401}\\
LiF & 8.882  & 8.874 & 14.20~\cite{PhysRevLett.102.226401}\\
MgO & 4.697 & 4.693 & 7.83~\cite{PhysRevLett.102.226401}\\
Ne & 11.496 & 11.448 & 21.7~\cite{PhysRevLett.102.226401}\\
\hline
MARE(\%) & 41.27 & 45.39 & 
\end{tabular}
\end{ruledtabular}
\caption{\label{tab:bangaps} Band gaps of several bulk materials, in eV, within different approximations and measured experimentally. }
\end{table}

\begin{table}
\begin{ruledtabular}
\begin{tabular}{lcccc}
Approx. & LDA$_{\rm xc}$ & LDA$_{\rm x}$+FBE$_{\rm c}^{\rm CS}$ & Exp.~\cite{PhysRevB.84.035117} \\
\hline
C & 3.539 & 3.540 & 3.553\\
Si & 5.394 & 5.404 & 5.421\\
GaAs & 5.723 & 5.733 & 5.640\\
LiF &  3.906 & 3.918 & 3.972\\
\hline
MARE(\%) & 1.006 & 0.922 & 
\end{tabular}
\end{ruledtabular}
\caption{\label{tab:lattice_constants} Lattice constants of several bulk materials, in \r{A}, within different approximations and measured experimentally. The mean averaged relative error (MARE) is also provided. }
\end{table}

\begin{table}
\begin{ruledtabular}
\begin{tabular}{lcccc}
Approx. & LDA$_{\rm xc}$ & LDA$_{\rm x}$+FBE$_{\rm c}^{\rm CS}$  & Exp.~\cite{PhysRevB.84.035117} \\
\hline
C & 461.99  & 459.48 & 454.7\\
Si & 95.30 &  93.59 & 100.8\\
GaAs & 79.95 & 83.036 & 76.7\\
LiF &  47.629 & 49.894 & 76.3 \\
\hline
MARE(\%) & 12.218 & 12.768 & 
\end{tabular}
\end{ruledtabular}
\caption{\label{tab:bulk_moduli} Bulk moduli of several bulk materials, in GPa, within different approximations and measured experimentally. }
\end{table} 

In order to assess the performance of the polarized version of the functional, we also investigated the performances of the correlation functional on elemental $3d$ ferromagnets. As discussed above, in case of spin-polarized systems, one option is to solve the Sturm--Liouville Eq.~\eqref{eq:ExchangeTimeDependent} for the spin-polarized correlation force of Eq.~\eqref{eq:correlation_force_pol}. Alternatively, one can use the approximated correlation potential from Eq.~\eqref{eq:polarized_potential_approx}. Table~\ref{tab:3d_metals} shows the local magnetic moment per atom of the different metals obtained by this method, together with the approximation of Eq.~\eqref{eq:polarized_potential_approx}. As expected, we found that this is a reasonable assumption, in agreement with prior work neglecting the same terms~\cite{PhysRevB.46.6671}, even if the results are found to be slightly worse than with the Sturm--Liouville equation. Overall, similar to the above results for semiconductors and insulators, we found that our correlation functional gives a similar performance as LSDA$_\mathrm{xc}$~\cite{perdew1981}.
\begin{table}
\begin{ruledtabular}
\begin{tabular}{lcccc}
Approx. & LSDA$_{\rm xc}$ & LSDA$_{\rm x}$+FBE$_{\rm c}^{\rm CS}$  &  LSDA$_{\rm x}$+FBE$_{\rm c}^{\rm CS}$ & Exp.\\
& & Eq.~\eqref{eq:correlation_force_pol} & Eq.~\eqref{eq:polarized_potential_approx} &  \\
\hline
Fe (bcc) & 2.305 & 2.258 & 2.251 & 1.98~\cite{PhysRevLett.75.152} \\
Ni (fcc) & 0.591 & 0.595 & 0.586 & 0.54~\cite{PhysRevB.43.6785}\\
Co (hcp) & 1.598 & 1.598 & 1.591 & 1.55~\cite{PhysRevLett.75.152}\\
\end{tabular}
\end{ruledtabular}
\caption{\label{tab:3d_metals} Local magnetic moments, in $\mu_{\rm B}$/atom,  within different approximations and experimentally measured spin magnetic moments. }
\end{table}  
\subsection{Finite systems}

Further, we performed benchmark calculations for isolated atoms. Here we did all-electron calculations for He, Be, B, and C using the Octopus code and a grid spacing of 0.1 Bohr, as well as Ne with 0.05 Bohr. The radius of the simulation box was taken to be 7, 8, 8, 9, and 10 Bohr for He, Be, B, C, and Ne, respectively. In Tab.~\ref{tab:atoms}, we compare the correlation energies obtained by the same approximations as used in the prior section in order to asses the performance of the correlation functional. 
Overall, we find that the correlation energy is better than obtained by a LDA correlation energy functional. We also compared our results with the ones from LDA exchange plus the related meta-GGA Colle--Salvetti correlation energy functional (LDA$_{\rm x}$+CS). As this meta-GGA functional retains second-order terms in the gradient expansion of the energy density, which would correspond to third-order term in the gradient expansion of the force density, the results are found to be much closer to the result inferred from experiments~\cite{savin1986application}. We expect that going beyond the first-order gradient expansion for the correlation force density would also improve the results for atoms, especially due to the dependence on the Laplacian of the density and on the kinetic energy density.
We also tested the influence of changing the exchange part from LDAx to FBEx, doing full exchange-correlation functional from the force-balance equation (FBE$_{\rm x}$ + FBE$_{\rm c}^{\rm CS}$). Overall, we find very little changes for the correlation energy due to the change of the exchange potential, as expected.
For all cases, the ionization potentials can be found in Tab.~\ref{tab:atoms_Ip} in the appendix.
\begin{table}
\begin{ruledtabular}
\begin{tabular}{lccccc}
Approx. & LDA$_{\rm xc}$  & LDA$_{\rm x}$ & LDA$_{\rm x}$ & FBE$_{\rm x}$  & Exp. \\
 & & +FBE$_{\rm c}^{\rm CS}$ & +CS &  +FBE$_{\rm c}^{\rm CS}$ & \cite{savin1986application} \\
\hline
He & -111 & -72  & -41  & -74  & -42  \\
Be & -223 & -150 & -93  & -152 & -94  \\
B  & -293 & -200 & -133 & -200 & -125 \\
C  & -370 & -257 & -178 & -257 & -156 \\
Ne & -737 & -547 & -375 & -552 & -387 \\
\hline
MARE(\%) & 132.70  & 59.42  & 5.41 & 61.05 & \\
\end{tabular}
\end{ruledtabular}
\caption{\label{tab:atoms} Atom correlation energies, in mHa, within different approximations and computed from experimental results (see text). }
\end{table}

\begin{figure}[ht]
  \begin{center}
  \includegraphics[width=\columnwidth]{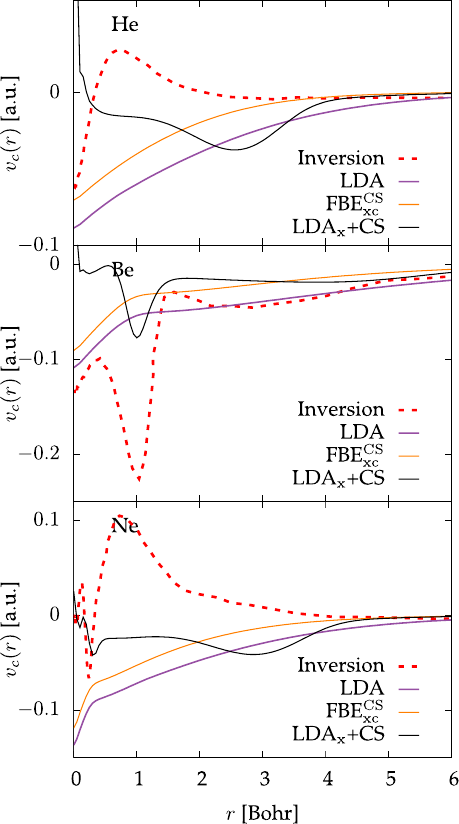}
    \caption{Correlation potential for different atoms and different functionals, see main text for details.}
    \label{fig:v_c}
  \end{center}
\end{figure}

The comparison with correlation potentials for He, Be, and Ne obtained from Kohn-Sham inversion of a many-body calculation for He~\cite{PhysRevA.50.3827}, and quantum Monte-Carlo simulations for Be and Ne~\cite{PhysRevA.54.4810}, are shown in Fig.~\ref{fig:v_c}.
While correlation energies  are improved compared to LDA, it is clear that the shape of the potential obtained by our proposed functional does not really change much when compared to LDA. The LDA$_{\rm x}$+CS results in the comparison show the relevance of higher-order terms in the gradient expansion.

% 
%%%%%%%%%%%%%%%%%%%%%%%%%%%%%%%%%%%%%%%%%%%%%%%%%%%%%%%%%%%%%%%%%%%%%%%%%%%%%%%%%%%%%%%%%%%%%%%%%%%%%%%%%%%%%%%%%%%%%%%%%%%%%%%%%
\section{Conclusions}
\label{sec:conclusions}
%%%%%%%%%%%%%%%%%%%%%%%%%%%%%%%%%%%%%%%%%%%%%%%%%%%%%%%%%%%%%%%%%%%%%%%%%%%%%%%%%%%%%%%%%%%%%%%%%%%%%%%%%%%%%%%%%%%%%%%%%%%%%%%%%%
%
Starting from an ansatz for the correlated wavefunction due to Colle and Salvetti~\cite{colle1975approximate}, we derived an analytical expression for the interaction exchange-correlation force density. By doing a short-range expansion of the two-body reduced density matrix, we obtained from the force density an exchange potential but also a correlation potential and the corresponding energies.
We have then compared the performance of our correlation functional to the standard LDA correlation energy functional based on the homogeneous electron gas, using the LDA exchange functional for the exchange part. We showed that our correlation functional compares favorably to the latter one for atoms, but is slightly worse for solids. 

With this we demonstrate that starting from the force-balance equation is a viable and interesting route towards the construction of analytical energy functionals that also include correlation effects. This method is particularly suited for incremental, analytical improvements.
Future work will address the derivation of the higher-order gradient expansion to get an improved correlation functional, possibly leading to a transverse part of the correlation vector field, which would allow for going beyond the adiabatic approximation in TDDFT, and to include memory effects.
We hope that this work will motivate further works on building functionals based on the force-balance equation.

\acknowledgments
This work was supported by the European Research Council (Grant No. ERC-2015-AdG694097), by the Cluster of Excellence
“CUI: Advanced Imaging of Matter” of the Deutsche Forschungs-gemeinschaft (DFG) – EXC 2056–Project ID 390715994, and the
Grupos Consolidados (Grant No. IT1249-19). 

\appendix

\section{Virial relation for the interaction force density}
\label{app:virial}

Here we derive the virial relation Eq.~\eqref{eq:virial} from the main text.
Since both spin components can be treated separately, the spin index is suppressed for conciseness.
We will first prove the virial relation for the complete Hxc energy,
\begin{equation}
 E_{\rm Hxc}^{\rm CS} = \int \d\rr\, \rr \cdot \FF_W^{\rm CS}(\rr).
\end{equation}
To derive this result, we use the symmetry of the expression in $\rr$ and $\rr'$ and the relation $ (\rr-\rr') \cdot(\nabla |\rr'-\rr|^{\alpha}) = \alpha |\rr'-\rr|^\alpha$ to get
\begin{equation}
\begin{aligned}
 \int &\d\rr\, \rr \cdot\FF_W(\rr) \\
 &= - 2\int \d\rr \int  \d\rr'\, \rr \cdot(\nabla |\rr'-\rr|^{-1}) \rho^{(2)}(\rr,\rr',\rr,\rr') \\
 &= - \int \d\rr \int  \d\rr'\, (\rr-\rr') \cdot(\nabla |\rr'-\rr|^{-1}) \rho^{(2)}(\rr,\rr',\rr,\rr') \\
 &= \int \d\rr \int \d\rr'\, \frac{\rho^{(2)}(\rr,\rr',\rr,\rr')}{|\rr'-\rr|}.
\end{aligned}
\end{equation}
Inserting the approximation 
$\rho^{(2)}(\rr,\rr',\rr,\rr') \approx \rho^{(2)}_{\rm s}(\rr,\rr',\rr,\rr') \left( 1 - 2\nu(\rr,\rr')  + \nu^2(\rr,\rr')\right)$ from Eq.~\eqref{eq:2rdm_CS} for the 2RDM, we recognize that we retrieve the sum of the Hartree energy, exchange energy, and the correlation energy of \citet[Eq.~(9)]{colle1975approximate}.
This shows that our force density is indeed the corresponding force density associated with the Colle--Salvetti correlation energy, at least before performing any gradient expansion.

\section{Derivation of the correlation force density}
\label{app:force_density}
We consider the vector integral of the form $\int\d\mathbf{s}\, f(s)F(\rr,\mathbf{s}) \mathbf{\hat{s}}$. The gradient expansion of the function $F$ leads to 
\begin{equation}
 F(\rr,\rr+\mathbf{s}) = F(\rr,\mathbf{0}) + \mathbf{s}\cdot[\nabla' F(\rr,\rr')]\Big|_{\rr'=\rr} + \mathcal{O}(s^2).
\end{equation}
The zero-order term in the integral vanishes and we consider in the following only the first order term. Going to the next non-zero order (third order) would lead to a semi-local meta-GGA functional. 
In order to compute the angular integral, we use the relation
\begin{equation}
 \int \d\Omega \, \mathbf{\hat{r}}(\mathbf{\hat{r}}\cdot\mathbf{a}) = \frac{4\pi}{3}\mathbf{a}
\end{equation}
for the surface integral, where $\mathbf{\, a}$ is a constant vector.
This leads to
\begin{eqnarray}
  \int \d\mathbf{s} \, f(s) \mathbf{\hat{s}} \left(\mathbf{s}\cdot[\nabla' F(\rr,\rr')]\Big|_{\rr'=\rr}\right) 
  %=  \int ds s f(s) \mathbf{\hat{s}} (\mathbf{\hat{s}}\cdot [\nabla' F(\rr,\rr')]\Big|_{\rr'=\rr})
  \nonumber\\
  = \frac{4\pi}{3}[\nabla' F(\rr,\rr')]\Big|_{\rr'=\rr}\int \d s\, f(s) s^3.
\end{eqnarray}
In our case we have
\begin{align}
 F&(\rr,\rr') = \rho^{(2)}_{\rm s}(\rr,\rr')\nonumber\\
 &\times \Bigg[ e^{-\beta^2(\rr)s^2}\left(1-\Theta\left(\frac{\rr+\rr'}{2}\right)\left(1+\frac{|\rr'-\rr|}{2} \right)\right)^2 \nonumber\\
&\qquad - 2\left(1-\Theta\left(\frac{\rr+\rr'}{2}\right)\left(1+\frac{|\rr'-\rr|}{2} \right)\right)\Bigg]
\end{align}
and $f(s) = \exp(-\beta^2(\rr)s^2)/s^2$.
The radial integral therefore gives 
\begin{equation}
 \int_0^\infty \d s\, f(s) s^3 = \int_0^\infty \d s\, s \,e^{-\beta^2(\rr)s^2} = \frac{1}{2\beta^2(\rr)}.
\end{equation}

\section{Numerical details and further results}
\label{app:parameters}

In this appendix we report the numerical parameters used to describe the different solids mentioned in Sec.~\ref{sec:results}. The parameters are given in Tab.~\ref{tab:param} for the bulk materials.
Ionization potentials for the studied atoms are also reported in Tab.~\ref{tab:atoms_Ip}.

\begin{table}
\begin{ruledtabular}
\begin{tabular}{lcccc}
System & Space group & $\Delta r$ & $k$-point grid & $a$ \\
& & [Bohr] & & [\AA] \\
\hline
AlAs & 216 & 0.40 & $6\times6\times6$ & 5.66\\
AlSb & 216 & 0.40 & $6\times6\times6$ & 6.14\\
GaAs & 216 & 0.25 & $10\times10\times10$ & 5.65325\\
ZnO & 186 & 0.25 & $8\times8\times8$ & 3.1995/5.1330\\
AlP & 216 & 0.35 & $8\times8\times8$ & 5.451\\
BN & 194 & 0.45 & $8\times8\times8$ & 2.503\\
TiO$_2$ & 136 & 0.25 & $8\times8\times8$ & 4.594/2.959 \\
Ar & 225 & 0.40 & $10\times10\times10$ & 5.256 \\
C & 227 & 0.25 & $8\times8\times8$ & 3.567\\
Si & 227 & 0.25  & $8\times8\times8$ & 5.431\\
LiF & 225 & 0.25 & $8\times8\times8$ & 4.026\\
MgO & 225 & 0.25 & $8\times8\times8$ & 4.212\\
Ne & 225 & 0.40 & $10\times10\times10$ & 4.429\\
Fe & 229 & 0.27 & $24\times24\times24$ & 2.867 \\ 
Ni & 255 & 0.27 & $24\times24\times24$ & 3.436 \\
Co & 194 & 0.27 & $24\times24\times24$ & 2.503/4.0574\\ 
\end{tabular}
\end{ruledtabular}
\caption{\label{tab:param} Numerical parameters used for bulk materials, see the text for details. The grid spacing $\Delta r$ and the lattice parameter $a$ are given in {\AA}ngström. For ZnO, TiO$_2$, and Co the two values correspond to the in-plane and out-of-plane lattice parameters.}
\end{table}

\begin{table}
\begin{ruledtabular}
\begin{tabular}{lccccc}
Approx. & He & Be & B & C & Ne\\
\hline
 LDA$_{\rm xc}$ & 15.506 & 5.559 & 3.650 & 6.348 & 13.594\\
LDA$_{\rm x}$+FBE$_{\rm c}^{\rm CS}$ & 14.993  &  5.007 & 3.167 & 6.165 & 13.081\\
LDA$_{\rm x}$+CS & 14.747  & 4.993 & 3.227 &  6.030 & 12.996\\
FBE$_{\rm x}$ +FBE$_{\rm c}^{\rm CS}$ & 26.035 &  9.028 & 3.747 &  5.870 & 24.483\\
Exp.~\cite{NIST_ASD} & 24.587 & 9.323 & 8.298 & 11.26 &  21.564 \\
\end{tabular}
\end{ruledtabular}
\caption{\label{tab:atoms_Ip} Ionization potential, in eV, within several approximations and experimental values, obtained from the highest occupied eigenvalue.}
\end{table}

%aipnum4-2.bst 2019-01-14 (MD) hand-edited version of apsrev4-1.bst
%Control: key (0)
%Control: author (8) initials jnrlst
%Control: editor formatted (1) identically to author
%Control: production of article title (0) allowed
%Control: page (1) range
%Control: year (1) truncated
%Control: production of eprint (0) enabled
%

\end{document}